\documentclass[prd,twocolumn,showpacs,nofootinbib,preprintnumbers]{revtex4}
\usepackage{amsmath,amsthm,amssymb}
\usepackage{amsfonts}
\usepackage{graphicx}
\usepackage{dcolumn}
\usepackage{hyperref}

\def\be{\begin{equation}}
\def\ee{\end{equation}}
\def\bea{\begin{eqnarray}}
\def\eea{\end{eqnarray}}

\def\5{\overline 5}

\begin{document}
\title{Coupled phantom field in loop quantum cosmology}

\author{Burin Gumjudpai}
\email{buring@nu.ac.th}   \affiliation{Fundamental Physics \&
Cosmology Research Unit \\ The Tah Poe Academia Institute (TPTP),
Department of Physics\\ Naresuan University, Phitsanulok, Thailand
65000}

\date{\today}

\vskip 1pc
\begin{abstract}
A model of phantom scalar field dark energy under exponential
potential coupling to barotropic dark matter fluid in loop quantum
cosmology is addressed here. We derive a closed-autonomous system
for cosmological dynamics in this scenario. The expansion in loop
quantum universe has a bounce even in presence of the phantom
field. The greater decaying from dark matter to dark phantom
energy results in greater energy storing in the phantom field.
This results in further turning point of the field. Greater
coupling also delays bouncing time. In the case of phantom
decaying, oscillation in phantom density makes small oscillation
in the increasing matter density.
\end{abstract}

\pacs{98.80.Cq}

\maketitle \vskip 1pc \vspace{1cm}

\section{Introduction}
There has recently been evidence of present accelerating expansion
of the universe from cosmic microwave background (CMB)
anisotropies, large scale galaxy surveys and type Ia supernovae
\cite{Spergel:2003cb, Scranton:2003in, Riess:1998cb}. Dark energy
(DE) in form of either cosmological constant or scalar field
matter is a candidate answer to the acceleration expansion which
could not be explained in the regime of standard big bang
cosmology \cite{Copeland:2006wr}. DE possesses equation of state
$p =w\rho$ with $w < -1/3$ enabling it to give repulsive gravity
and therefore accelerate the universe. Combination of
observational data analysis of CMB, Hubble Space Telescope, type
Ia Supernovae and 2dF datasets allows constant $w$ value between
-1.38 and -0.82 at the 95 \% of confident level
\cite{Melchiorri:2002ux}. Meanwhile, assuming flat universe, the
analysis result, $-1.06 < w < -0.90$ has been reported by
\cite{Spergel:2006hy} using WMAP three-year results combined with
Supernova Legacy Survey (SNLS) data. Without assumption of flat
universe, mean value of $w$ is -1.06 (within a range of -1.14 to
-0.93). Most recent data (flat geometry assumption) from ESSENCE
Supernova Survey Ia combined with SuperNova Legacy Survey Ia gives
a constraint of $w = -1.07 \pm 0.09$ \cite{Wood-Vasey:2007jb}.
Observations above show a possibility that a fluid with $w < -1$
could be allowed in the universe \cite{Caldwell:1999ew}. This type
of cosmological fluid is dubbed {\it phantom}. Conventionally
Phantom behavior arises from having negative kinetic energy term.

Dynamical properties of the phantom field in the standard FRW
cosmology were studied before. However the scenario encounters
singularity problems at late time
\cite{Li:2003ft,Urena-Lopez:2005zd}. While investigation of
phantom in standard cosmological model is still ongoing, there is
an alternative approach in order to resolve the singularity
problem by considering phantom field evolving in Loop Quantum
Cosmology (LQC) background instead of standard general
relativistic background \cite{Samart:2007xz, Naskar:2007dn}. Loop
Quantum Gravity-LQG is a non-perturbative type of quantization of
gravity and is background-independent
\cite{Thiemann:2002nj,Ashtekar:2003hd}. LQG provides cosmological
background evolution for LQC. An effect from loop quantum
modification gives an extra correction term $-{\rho^2}/\rho_{\rm
lc}$ into the standard Friedmann equation
\cite{Bojowald:2001ep,Date:2004zd, Singh:2006sg}. Problem for
standard cosmology in domination of phantom field is that it leads
to singularity, so called the Big Rip \cite{Caldwell:2003vq}. The
$-{\rho^2}/\rho_{\rm lc}$ term, when dominant at late time, causes
bouncing of expansion hence solving Big Rip singularity problem
\cite{Ashtekar:2003hd, Bojowald:2001xe, Ashtekar:2006rx}.
Recently, a general dynamics of scalar field including phantom
scalar field coupled to barotropic fluid has been investigated in
standard cosmological background. In this scenario, the scaling
solution of the coupled phantom field is always unstable and it
can not yield the observed value $
 \Omega_\phi \sim 0.7 $ \cite{Gumjudpai:2005ry}.
Indeed there should be other effects from loop quantum correction
to the Friedmann equation. Moreover when including potential term
in scalar field density, the quantum modification must be included
\cite{Bojowald:2006gr}. Although, the Friedmann background is
valid only in absence of field potential, however, investigation
of a phantom field evolving under a potential could reveal some
interesting features of the model. In this letter, we investigate
a case of coupled phantom field in LQC background in alternative
to the standard relativistic cosmology case. In Section
\ref{sec2}, we introduce framework of cosmological equations
before considering dynamical autonomous equations in Section
\ref{sec3}. We show some numerical results in Section \ref{sec4}
where the coupling strength is adjusted and compared. Conclusion
and comments are in Section \ref{sec5}.
\section{COSMOLOGICAL EQUATIONS} \label{sec2}
\subsection{Loop quantum cosmology}
The effective flat universe Friedmann equation from LQC is given
as \cite{Bojowald:2001ep, Singh:2006sg},
\begin{eqnarray}
H^2&=&\frac{\rho}{3M_{\rm P}^2}\left(1-\frac{\rho}{\rho_{\rm
lc}}\right), \label{fr}
\end{eqnarray}
where $H$ is Hubble constant,   $M_{\rm P}$ is reduced Planck
mass, $\rho$ is density of cosmic fluid,  $\rho_{\rm lc} =
\sqrt{3}/(16 \pi \zeta^3 G^2 \hbar)$. The parameter $\zeta$   is
Barbero-Immirzi dimensionless parameter and $G$ is the Newton's
gravitational constant.
\subsection{Phantom scalar field}
Nature of the phantom field can be extracted from action,
\begin{eqnarray}
S = \int {\rm d}^4x \sqrt{-g}\left[\frac{1}{2}(\partial^a\phi)(
\partial_a\phi) - V(\phi) \right]. \label{l}
\end{eqnarray}
Energy density and pressure are given by
\begin{eqnarray}
\rho_{\phi} = -\frac{1}{2}\dot{\phi}^2 + V(\phi),
\end{eqnarray}
and
\begin{eqnarray}
p_{\phi} = -\frac{1}{2}\dot{\phi}^2 - V(\phi),
\end{eqnarray}
with equation of state,
\begin{eqnarray}
w_{\phi} \equiv \frac{p_{\phi} }{\rho_{\phi}} = \frac{\dot{\phi}^2
+ 2V(\phi)}{\dot{\phi}^2 - 2V(\phi)}.
\end{eqnarray}
When the field is slowly rolling, the approximate value of $w$ is
-1. As long as the approximation, $\dot{\phi}^2 \sim 0 $ or the
condition, $\dot{\phi}^2 < 2V $ holds, $w$  is always less than
-1. In our scenario, the universe contains two fluid components.
These are barotropic fluid with equation of state $p_{\rm m} =
\rho_{\rm m}w_{\rm m}$ and phantom scalar field fluid. The total
density is then $\rho = \rho_{\rm m} + \rho_{\phi}$ which governs
total dynamics of the universe.
\subsection{Coupled phantom scalar field}
Here we consider both components coupling to each other. Fluid
equations for coupled scalar fields proposed by
\cite{Piazza:2004df} assuming flat standard FRW universe are
\begin{eqnarray}
\dot{\rho}_{\phi} + 3H(1+w_{\phi})\rho_{\phi} &=& -Q\rho_{\rm
m}\dot{\phi}\,, \label{coup1} \\
\dot{\rho}_{\rm m} + 3H(1+w_{\rm m})\rho_{\rm m} &=& +Q\rho_{\rm
m}\dot{\phi}\,. \label{coup2}
\end{eqnarray}
These fluid equations contain a constant coupling between dark
matter (the barotropic fluid) and dark energy (the phantom scalar
field) as in \cite{Amendola:1999er}. Eqs. (\ref{coup1}) and
(\ref{coup2}) can also be assumed as conservation equations of
fluids in the LQC. Total action for matter and phantom scalar
field is \cite{Piazza:2004df}
\begin{eqnarray}
S = \int {\rm d}^4x \sqrt{-g}\left[ \frac{M_{\rm P}^2}{2}R +
p(X,\phi) \right] + S_{\rm m}(\phi)\,. \label{actionmat}
\end{eqnarray}
Assuming scaling solution of the dark energy in standard
cosmology, the pressure Lagrangian density is written as
\begin{eqnarray}
p(X,\phi) = -X -c \exp(-\lambda\phi/M_{\rm P}^2)\,,\label{L}
\end{eqnarray}
where $X$ is the kinetic term,
$-g^{ab}\partial_a\phi\partial_b\phi/2$ of the Lagrangian density
(\ref{l}) and (\ref{L}). The second term on the right of Eq.
(\ref{L}) is exponential potential, $V(\phi) = c
\exp(-\lambda\phi/M_{\rm P}^2)$ which gives scaling solution for
canonical and phantom ordinary scalar field in standard general
relativistic cosmology when steepness of the potential, $\lambda$
is fine tuned as
\begin{eqnarray}
\lambda = Q \frac{1 + w_{\rm m} - \Omega_{\rm \phi}(w_{\rm
m}-w_{\rm \phi})}{\Omega_{\rm \phi}(w_{\rm m}-w_{\rm \phi})}\,.
\label{lambda}
\end{eqnarray}
The steepness (\ref{lambda}) is, in standard cosmological
circumstance, constant in the scaling regime due to constancy of
$w_{\phi}$ and $\Omega_{\phi}$ \cite{Piazza:2004df,
Gumjudpai:2005ry}. However, in LQC case, there has been a report
recently that the scaling solution does not exist for phantom
field evolving in LQC \cite{Samart:2007xz}. Therefore our spirit
to consider constant $\lambda$ is the same as in
\cite{Copeland:1997et} not a motivation from scaling solution as
in \cite{Piazza:2004df}. The exponential potential is also
originated from fundamental physics theories such as higher-order
gravity \cite{highergrav} or higher dimensional gravity
\cite{string}.
\section{COSMOLOGICAL DYNAMICS} \label{sec3}
\begin{figure}[t]
\begin{center}
\includegraphics[width=7.5cm,height=7.0cm,angle=0]{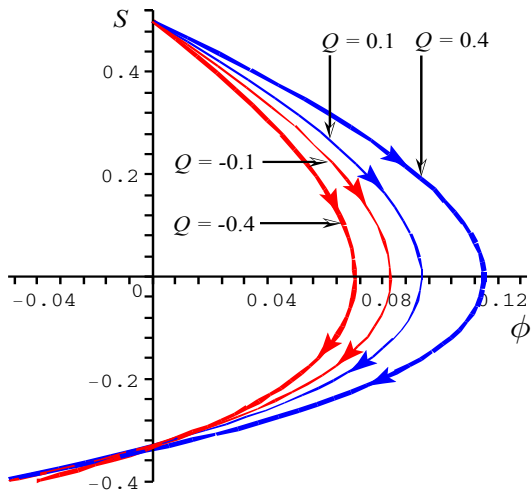}
\end{center}
\caption{Phase portrait of  $ S(t)$  versus  $ \phi(t) $ for  $Q
 = -0.4, -0.1, 0.1$ and $0.4$ from left to right. All trajectories
have the same initial conditions  $S(0)=0.5 $ and  $ \phi(0)=0$. }
\label{phase}
\end{figure}
\begin{figure}[t]
\begin{center}
\includegraphics[width=8.0cm,height=7.0cm,angle=0]{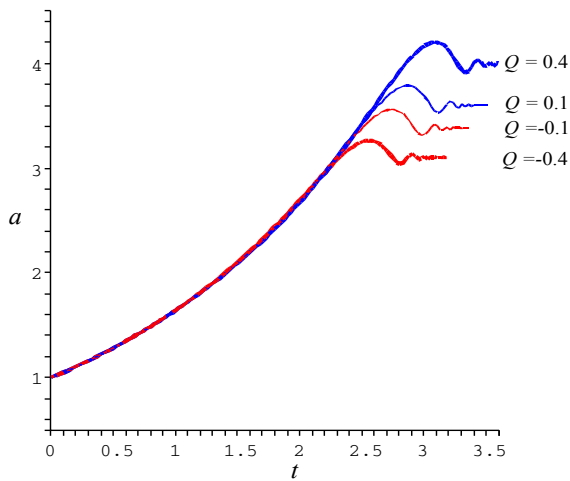}
\end{center}
\caption{Scale factor plotted versus time for $Q = -0.4, -0.1,
0.1$ and $0.4$ (from bottom to top). } \label{scale}
\end{figure}
Time derivative of the effective LQC Friedmann equation LQC
(\ref{fr}) is
\begin{eqnarray}
\dot{H} &=& - \frac{(\rho + p)}{2 M_{\rm P}^2}\left(1 -
\frac{2\rho}{\rho_{\rm lc}}\right), \label{dh1}\\
&=& - \frac{(1+w_{\phi})\rho_{\phi} + (1+w_{\rm m})\rho_{\rm m}}{2
M_{\rm P}^2} \left[ 1 - \frac{2}{\rho_{\rm lc}}(\rho_{\phi}+
\rho_{\rm m}) \right], \label{dh2} \nonumber\\ \\
&=& - \frac{\left[-S^2+(1+w_{\rm m})\rho_{\rm m}\right] }{2 M_{\rm P}^2}\times \nonumber \\
& &\left[ 1 -\frac{2}{\rho_{\rm lc}}\left( - \frac{S^2}{2} + c
e^{-\lambda\phi/M_{\rm P}^2}+ \rho_{\rm m}\right)\right].
\label{1}
\end{eqnarray}
In above equations we define new variable
\begin{eqnarray}
S \equiv \dot{\phi}. \label{2}
\end{eqnarray}
The coupled fluid equations (\ref{coup1}) and (\ref{coup2}) are
re-expressed in term of $S$ as
\begin{eqnarray}
\dot{S} &=& -3HS + \frac{{\rm d}V}{{\rm d}\phi} + Q\rho_{\rm m}\,, \label{3}\\
\dot{\rho}_{\rm m} &=& -3H(1+w_{\rm m})\rho_{\rm m} +Q\rho_{\rm
m}S\,. \label{4}
\end{eqnarray}
The Eqs. (\ref{1}), (\ref{2}), (\ref{3}) and (\ref{4}) form a
closed autonomous set of four equations. The variables here are
$\rho_{\rm m}$, $S$, $\phi$ and $H$. The autonomous set recovers
standard general relativistic cosmology in the limit $\rho_{\rm
lc}\rightarrow\infty $. The general relativistic limit affects
only the equation involving $H$. From the above autonomous set,
one can do a qualitative analysis with numerical integration
similar to phase plane presented in different situation
\cite{Gumjudpai:2003vv}. Another approach of analysis is to
consider a quantitative analysis \cite{Gumjudpai2007}.
\begin{figure}[t]
\begin{center}
\includegraphics[width=8.0cm,height=6.2cm,angle=0]{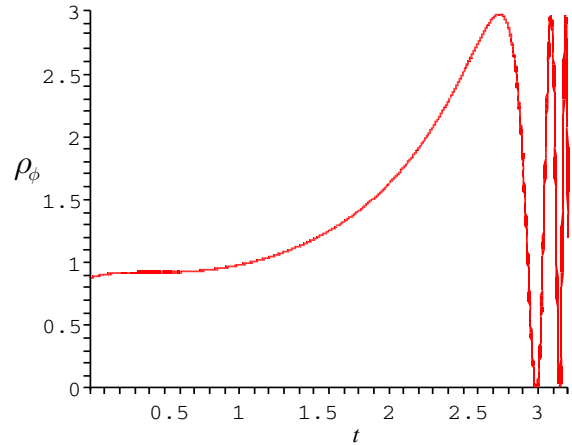}
\end{center}
\caption{Phantom field density plotted versus time for $Q = -0.1$.
The other values of $Q$ also yield bouncing and oscillation.}
\label{rhoQ}
\end{figure}
\section{NUMERICAL SOLUTIONS} \label{sec4}
Here we present some numerical solution for a positive and
negative coupling between the phantom field and barotropic fluid.
The solutions presented here are physically valid solutions
corresponding to Class II solutions characterized in
\cite{Samart:2007xz}. For non-minimally coupled scalar field in
Einstein frame \cite{Uzan:1999ch}, the coupling $Q$ lies in a
range $-1/\sqrt{6} < Q<1/\sqrt{6} $. Here we set  $Q = -0.4, -0.1,
0.1 $ and $0.4 $ which lie in the range. Effect of the coupling
can be seen from Eqs. (\ref{coup1}) and (\ref{coup2}). Negative
$Q$ enhances decay rate of scalar field to matter while giving
higher matter creation rate. On the other hand, positive $Q$
yields opposite result. Greater magnitude of $Q < 0$ gives higher
decay rate of the field to matter. Greater magnitude of $Q > 0$
will result in higher production rate of phantom field from
matter.
\subsection{Phase portrait}
The greater $Q$ value results in greater value of the field
turning point (see $\phi$-intercept in Fig. \ref{phase}). The
kinetic term $S(t)$ turns negative at the turning points
corresponding to the field rolling down and then halting before
rolling up the hill of exponential potential. When $Q$ is greater,
the field can fall down further, hence gaining more total energy.
The result agrees with the prediction of Eqs. (\ref{coup1}) and
(\ref{coup2}).
\subsection{Scale factor}
From Fig. \ref{scale}, the bounce in scale factor occurs later for
greater $Q$ value of which the phantom field production rate is
higher. The field has more phantom energy to accelerate the
universe in counteracting the effect of loop quantum (the bounce).
For less positive $Q$, the phantom production rate is smaller, and
for negative $Q$, the phantom decays. Therefore it has less energy
for accelerating the expansion in counteracting with the loop
quantum effect. This makes the bounce occurs sooner.
\subsection{Energy density}
\begin{figure}[t]
\begin{center}
\includegraphics[width=8.0cm,height=6.6cm,angle=0]{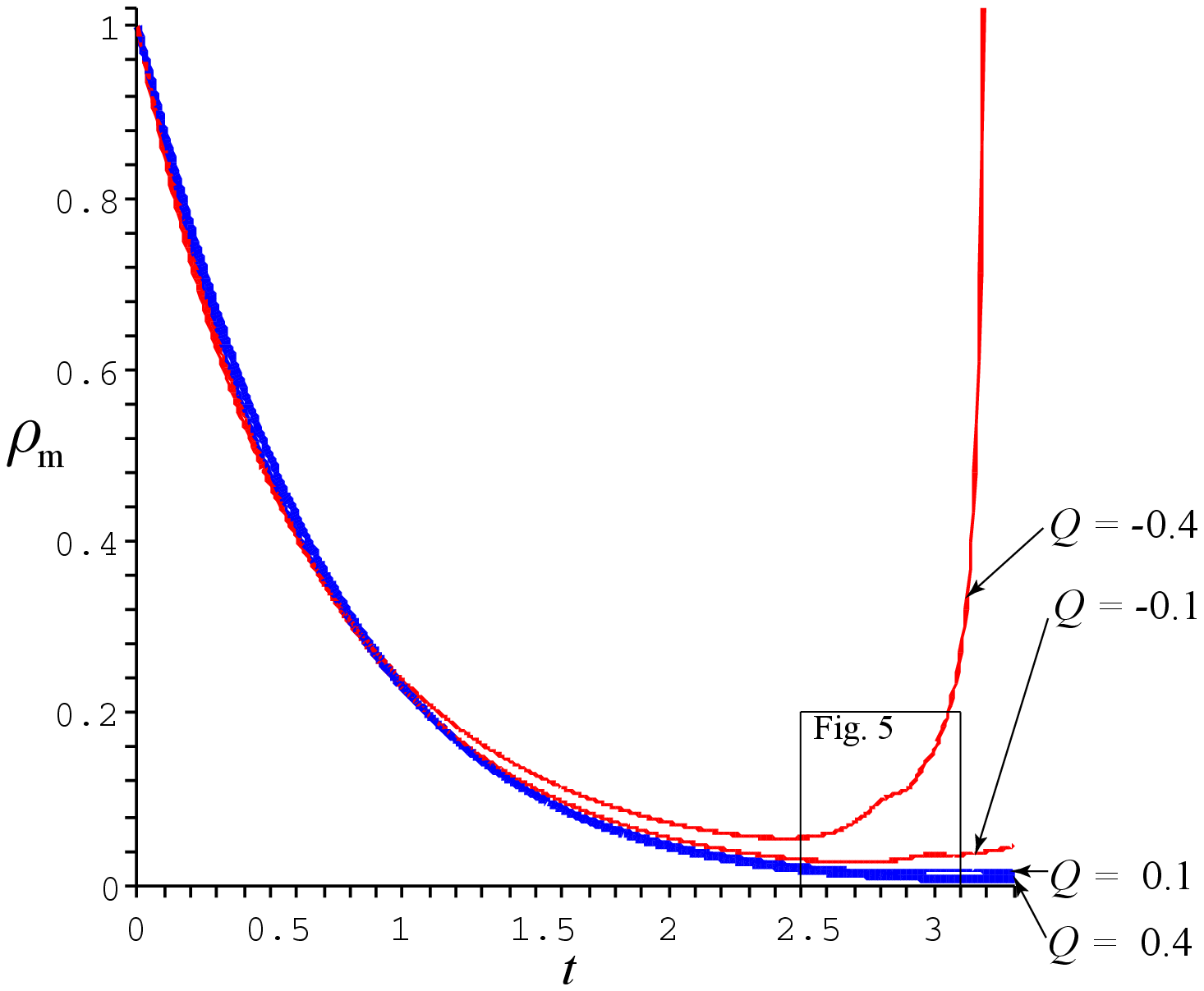}
\end{center}
\caption{Matter density plotted versus time for $Q  = -0.4, -0.1,
0.1$ and $0.4$ (from top to bottom).} \label{rhom}
\begin{center}
\includegraphics[width=8.0cm,height=7.0cm,angle=0]{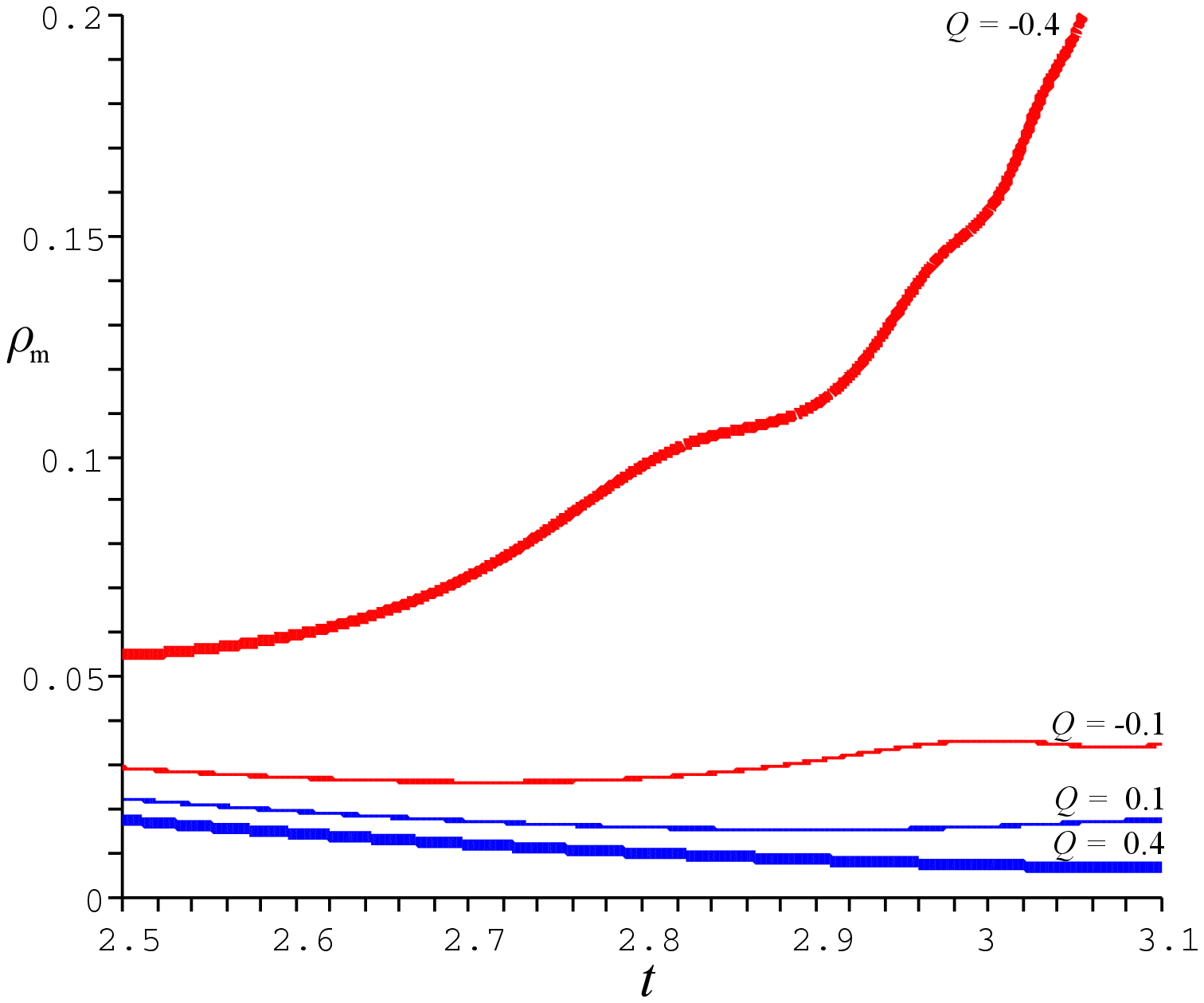}
\end{center}
\caption{Zoom-in portion of Fig. \ref{rhom}. The phantom field
decays to matter at highest rate for $ Q  = -0.4 $ (top line).
Oscillation in matter density due to oscillation in the phantom
field density is seen clearly here. } \label{rhomzoom}
\end{figure}

Time evolutions of energy density of the matter and the phantom
field are presented in Figs. \ref{rhoQ} and \ref{rhom}. If $Q >
0$, the matter decays to phantom. This reduces density of matter.
While for $Q < 0$, the matter gains its density from decaying of
phantom field. In Fig. \ref{rhoQ} there is a bounce of phantom
density before undergoing oscillation. For a non-coupled case, it
has recently been reported that the phantom density also undergoes
oscillation \cite{Samart:2007xz}. As seen in Figs. \ref{rhom} and
\ref{rhomzoom}, the oscillation in phantom density of the phantom
decay case ($Q < 0$) affects in oscillation in matter density
while for the case of matter decay ($Q > 0$), the matter density
is reduced for stronger coupling. The oscillation in the phantom
density comes from oscillation of the kinetic term $\dot{\phi}$,
i.e. $S(t)$ as shown in \cite{Samart:2007xz}.
\section{Conclusion and comments} \label{sec5}
In this letter, we have derived an autonomous system of a loop
quantum cosmological equations in presence of phantom scalar field
coupling to barotropic matter fluid. We choose constant coupling
$Q$ between matter and the phantom field to positive and negative
values and check numerically the effect of $Q $ values on (1)
phase portrait, (2) scale factor and (3) energy density of phantom
field and matter. We found that field value tends to roll up the
hill of potential due to phantom nature. With greater $Q$, the
field can fall down on the potential further. This increases total
energy of the field. For canonical scalar field either standard or
phantom, LQC yields a bounce. The bounce is useful since it is
able to avoid Big Bang singularity in the early universe. Here our
numerical result shows a bouncing in scale factor at late time.
This is a Type I singularity avoidance even in presence of phantom
energy. The greater coupling results in more and more phantom
density. Greater phantom effect therefore delays the bounce, which
is LQC effect, to later time. In the case of matter decay to
phantom ($Q > 0$), oscillation in phantom energy density does not
affect matter density. On the other hand, when $Q < 0$, phantom
decays to matter, oscillation in phantom density results in
oscillation in the increasing matter density.

This work considers only the effects of sign and magnitude of the
coupling constant to qualitative dynamics and evolution of the
system. Studies of field dependent effects of coupling $Q(\phi)$
in some scalar-tensor theory of gravity and investigation of an
evolution of effective equation of state could also yield further
interesting features of the model. Quantitative dynamical analysis
of the model under different types of potential is also motivated
for future work. Frequency function of the oscillation in scale
factor and phantom density are still unknown in coupled case. It
looks like that the oscillation frequency tends to increase. This
could lead to infinite frequency of oscillation which is another
new singularity.

\vspace{0.5cm}

{\bf Acknowledgements:} B.~G. is grateful to Shinji Tsujikawa for
discussion. This work was presented as an invited talk at the SIAM
Physics Congress 2007. B.~G. thanks Thiraphat Vilaithong and the
committee of the Thai Physics Society for an invitation to present
this work at the congress. This work is supported by a TRF-CHE
Research Career Development Grant of the Thailand Research Fund
and the Royal Thai Government's Commission on Higher Education.

\end{document}